\documentclass[11pt]{article}
\usepackage[a4paper,bindingoffset=0.2in,%
            left=0.9in,right=0.9in,top=1in,bottom=1in,%
            footskip=.25in]{geometry}
\usepackage{color,soul}
\usepackage{amsmath,graphicx}
\usepackage{amssymb}
\usepackage{amsfonts}
\usepackage{amsmath, nccmath}
\usepackage{tipa}
\usepackage{mathtools}   
\DeclarePairedDelimiter{\abs}{\lvert}{\rvert}
\DeclarePairedDelimiter{\norm}{\lVert}{\rVert}
\usepackage{amsthm}
\theoremstyle{plain}
\newtheorem{theorem}{Theorem}
\newtheorem{corollary}[theorem]{Corollary}
\newtheorem{lemma}[theorem]{Lemma}
\newtheorem{proposition}[theorem]{Proposition}
\theoremstyle{definition}
\newtheorem{definition}[theorem]{Definition}

\theoremstyle{remark}
\newtheorem{remark}[theorem]{Remark}

\providecommand{\keywords}[1]{\textbf{\textit{Index terms---}} #1}
\usepackage{acronym}
\acrodef{CS}{Compressed Sensing}
\acrodef{BP}{Basis Pursuit}
\acrodef{RIP}{Restricted Isometry Property}
\acrodef{BOS}{Bounded Orthonormal System}

\title{Sparse recovery in Wigner-D basis expansion}
\author{Arya Bangun, Arash Behboodi and Rudolf Mathar
}
\date{}

\begin{document}
\maketitle
\begin{abstract}
{We are concerned with the recovery of $s-$sparse Wigner-D expansions in terms of $N$ Wigner-D functions. Considered as a generalization of spherical harmonics, Wigner-D functions are eigenfunctions of Laplace-Beltrami operator and form an orthonormal system. However, since they are not uniformly bounded, the existing results on \ac{BOS} do not apply. Using previously introduced preconditioning technique, a new orthonormal and bounded system is obtained for which \ac{RIP} property can be established. 
We show that the number of sufficient samples for sparse recovery scales with ${N}^{1/6} \,s\, \log^3(s) \,\log(N)$. 
The  phase transition diagram for this problem is also presented. We will also discuss the application of our results in the spherical near-field antenna measurement.}
\end{abstract}
\section{Introduction}
Consider a general function expanded in terms of $N$ orthonormal basis functions. If the expansion is sparse, in other words if it contains only a few non-zero coefficients, it is of high practical interest to see whether the coefficients can be recovered by a number of samples smaller than $N$. In general, \ac{CS} theory is concerned with the recovery of sparse signals using only few samples and it has sparkled a significant amount of research after the pioneering works in \cite{candes2006robust, donoho_compressed_2006}. 
It aims at finding necessary and sufficient conditions for signal recovery by a low complexity solving algorithm, and also discusses the stability of the solution to measurement noise and its robustness to sparsity defect. One recurring method is \ac{BP} algorithm which consists of minimizing the $\ell_1$-norm. A well known sufficient condition for recovery using \ac{BP}, as well as  other algorithms such as greedy methods and thresholding algorithms, is the \ac{RIP} property of the measurement matrix \cite{foucart2013mathematical}. Every matrix satisfying \ac{RIP} property of order $s$ is intuitively almost a norm-preserving transformation on the space of $s-$sparse vectors. \ac{RIP} has been first introduced in \cite{candes2005decoding} and later it has been shown that  Gaussian random matrices satisfy it \cite{candes_near-optimal_2006}. The stability and robustness of \ac{BP} under the \ac{RIP} condition is discussed in \cite{candes2006robust}. In general, subgaussian random matrices satisfy the \ac{RIP} and can be used for recovery \cite{foucart2013mathematical}.

However in most of the practical scenarios of interest and particularly in case of orthonormal systems, the measurement matrix possesses particular structures and complete random measurements such as Gaussian is not possible. For instance, the rows of the measurement matrix in an orthonormal system, are samples of $N$ orthonormal functions at a given time. If $m$ sampling times are chosen randomly and independently according to certain measure, the $m \times N$ resulting matrix can be shown to satisfy the \ac{RIP} property for \ac{BOS}. 
First it was shown in \cite{candes_near-optimal_2006} that partial random Fourier matrices satisfy \ac{RIP} with $\Omega(s\log^6(N))$ measurements. After further improvements in \cite{rudelson_sparse_2008}, the authors in \cite{cheraghchi_restricted_2013} manage to improve the bound on measurement numbers for \ac{BOS} to  $\Omega(s\log^3(s)\log(N))$. Recently Bourgain tightened the bound  for Hadamard matrices to $\Omega(s\log(s)\log^2(N))$ \cite{klartag_improved_2014} and similar results has been obtained in \cite{nelson2014new}. For discrete Fourier transform matrix, the bound has been improved to $\Omega(s\log(N))$ \cite{bandeira_discrete_2015} for $s$ dividing $N$.

For the existing bounds, the number of necessary measurements $m$ is scaled with $K^2$, where $K$ is the bound on basis functions. Therefore the bound on basis functions should be either independent of $N$ or should scale with lower powers of $N$.  A trivial lower bound for $K$ is equal to one and it is obtained for instance by Fourier matrix. Legendre polynomials, defined on $[-1,1]$, are bounded by $K=\sqrt{2N-1}$ which provides useless bound on $m$. However if the inner product of functions is taken according to Chebyshev probability measure and Legendre polynomials are weighted accordingly to guarantee orthogonality, the bound is found to be $K =\sqrt{3}$ \cite{rauhut2012sparse}. The change of measure serves to damp Legendre functions at the endpoints of the interval where they have the biggest growth. The technique is used later on for bounding spherical harmonics. Spherical harmonics, functions of azimuth $\phi$ and elevation $\theta$, are basis functions for the Hilbert space of $L^2$ functions on  $\mathbb{S}^2$. The spherical harmonics are bounded by $K \approx N^{1/2}$, and therefore produce useless bound on the required measurements. Using similar change of measure technique, Rauhut and Ward improved the bound to $K\approx N^{1/8}$ with preconditioning each function by $\sin(\theta)^{1/2}$ and using the product measure on sphere instead of the uniform measure\cite{rauhut2011sparse}. Using the measure $|\tan(\theta)|^{1/3}\rm{d}\theta\rm{d}\phi$, the bound has been later improved in  \cite{burq2012weighted} to  $K\approx N^{1/{12}}$ which applies, in a more general setting, to joint eigenfunctions of the Laplace-Beltrami operator on arbitrary surfaces of revolution. Combined with bounds on \ac{BOS}, the required measurements for recovery turns out to be $\Omega(N^{1/6}s\log^3(s)\log(N))$. 

In this work, similar to the works in \cite{rauhut2011sparse,burq2012weighted}, we derive a sufficient condition for recovery of sparse coefficients of Wigner-D expansions \cite{wigner2012group}. Considered as an extension of spherical harmonics, Wigner-D functions are used widely in near-field antenna measurements as the basis of linear equation of the electromagnetic fields the antenna. Although an eigenfunction of the Laplace-Beltrami operator, the Wigner-D functions are not defined on $\mathbb{S}^2$ but rather on the group of all rotations in $\mathbb{R}^3$, namely $\mathrm{SO}(3)$ which is a compact three dimensional manifold. The expansion in terms of $N$ Wigner-D functions involves Jacobi polynomials of lesser degree and therefore provides better bounds after preconditioning. The main steps consist of proving \ac{RIP} for random samples of preconditioned functions and then applying recovery results from \ac{BOS}. 

The paper is organized as follows. In section \ref{sec:Background}, the main definitions are provided as well as the main theorems used throughout the paper. In section \ref{sec:Main}, the \ac{RIP} theorem for Wigner-D functions and the theorem on sufficient conditions for recovery are stated. 
The numerical experiments and the application in spherical near-field measurement are presented in section \ref{sec:Num}. {Finally, The proof of main results is relegated to the appendix.}
\subsection{Notation}
$\theta$ is used for elevation and $\phi$ for azimuth. $\mathbb{N}$ is the set of natural numbers including zero. Vectors are presented by small bold letters and matrices by capital bold letters. the set $\{1,...,N\} $ is denoted by $[N]$.
\section{Definitions and Background}
\label{sec:Background}
\subsection{Wigner-D Functions}

Wigner-D functions form an orthogonal basis for the group of all rotations on 3-dimensional space  ${{\mathrm{SO}}(3)}$ \cite{wigner2012group}.
\begin{definition}
Wigner-D function of degree $l$ and orders $k$ and $n$ are defined as follows:
\begin{equation}
\mathrm{D}_l^{k,n}(\theta,\phi,\chi)= N_l e^{-jk\phi} \mathrm{d}_l^{k,n}(\cos \theta)  e^{-jn\chi} \label{def:WigD}
\end{equation}
where   $\theta \in [0,\pi]$, $\phi\in [0,2\pi)$ and $\chi\in [0,2\pi)$ are  Euler angles and $N_l=\sqrt{\frac{2l+1}{8\pi^2}}$ is a normalization factor to get the orthonormal property. Here $\mathrm{d}_l^{k,n}(\cos \theta)$ is Wigner-d function and is defined as:
\begin{equation}
\mathrm{d}_l^{k,n}(\cos \theta)= \omega \sqrt{\gamma} \sin^{\mu} \bigg(\frac{\theta}{2}\bigg)\cos^{\lambda}\bigg(\frac{\theta}{2}\bigg) P_{\alpha}^{(\mu,\lambda)}(\cos \theta)
\end{equation}
where $\gamma=\frac{\alpha!(\alpha + \mu + \lambda)!}{(\alpha+\mu)!(\alpha+\lambda)!}$, $\mu=\abs{k-n}$, $\lambda=\abs{k+n}$, $\alpha=l-\big(\frac{\mu+\lambda}{2}\big)$ and 
\begin{equation*}
 \omega= \begin{cases*} 
        1  & if $n\geq k $ \\
        (-1)^{n-k} & if $n<k$ 
                \end{cases*}
\end{equation*}
with degree $0 \leq l \leq \infty $ and order $-l \leq k,n \leq l$, $\forall l,k,n \in \mathbb{N}$. The function $P_{\alpha}^{(\mu,\lambda)}$ is the Jacobi polynomial. 
\label{def:1}
\end{definition}
Note that the Wigner-D function can be considered as the product of three functions, each one belonging to different set of orthonormal functions. Particularly, Wigner-d functions are weighted Jacobi polynomials that are orthogonal with respect to normalized Lebesgue measure. 
\begin{remark}
Wigner-D functions are sometimes called generalized spherical harmonics. The relation between Wigner-D function $\mathrm{D}_l^{k,n}$ and spherical harmonics $Y_{l}^k$ is as follows 
\begin{equation}\label{gener_SH}
\mathrm{D}_l^{-k,0}(\theta,\phi,0) = (-1)^k \sqrt{\frac{4\pi}{2l+1}}Y_{l}^k(\theta,\phi).
\end{equation}
Wigner-d function on the other hand is related to associated Legendre polynomials as follows:
\begin{equation}
\mathrm{d}_l^{k,0}(\cos \theta)= \sqrt{\frac{(l-k)!}{(l+k)!}}P_l^k(\cos \theta)
\end{equation}
\end{remark}

\begin{definition}[Wigner-D expansion]
The expansion of the function $g \in L^2({\mathrm{SO}(3)})$ in terms of Wigner-D functions $\mathrm{D}_l^{k,n}(\theta,\phi,\chi)$ writes as 
\begin{equation}
g(\theta,\phi,\chi)=\sum_{l=0}^{\infty}\sum_{k=-l}^{l}\sum_{n=-l}^{l} \hat{g}_l^{k,n} \,\mathrm{D}_l^{k,n}(\theta,\phi,\chi).
\label{wfexp}
\end{equation} 
This is also called the ${\mathrm{SO}(3)}$ Fourier expansion with Fourier coefficient $\hat{g}_l^{k,n}$ where
\begin{equation}
\hat{g}_l^{k,n}=\int_{0}^{2\pi}\int_{0}^{2\pi}\int_{0}^{\pi} g(\theta,\phi,\chi) \,\overline{\mathrm{D}_l^{k,n}(\theta,\phi,\chi)} \sin\theta \mathrm{d}\theta \mathrm{d}\phi \mathrm{d}\chi.
\end{equation} 
\end{definition}
Wigner-D functions are orthonormal with respect to the uniform measure on the sphere  $\mathrm{d}\nu=\sin \theta \mathrm{d}\theta \mathrm{d}\phi \mathrm{d}\chi$, namely:
\begin{equation}
\begin{aligned}
& \int_{0}^{2\pi}\int_{0}^{2\pi}\int_{0}^{\pi} \mathrm D_l^{k,n}(\theta,\phi,\chi) \overline{\mathrm D_{l'}^{k',n'} (\theta,\phi,\chi)} \sin \theta \mathrm{d}\theta \mathrm{d}\phi \mathrm{d}\chi \quad = \delta_{ll'} \delta_{kk'} \delta_{nn'}
\end{aligned}
\end{equation}
where $\delta_{ll'}$ is Kronecker delta. In this work, instead of infinite  expansion, we suppose that the functions are bandlimited. A function $g \in L^2(\mathrm{SO}(3))$ is bandlimited with bandwidth $B$ if it is expressed in terms of Wigner-D functions of degree less than $B$. A bandlimited function is said to be $s-$sparse if the vector of Wigner-D coefficients, $\mathbf{g}=(\hat{g}_l^{k,n})$ for  $0 \leq l \leq B-1, -l \leq k,n \leq l$ , is $s$-sparse, that is $\norm{ \mathbf g}_0\leq s$. Alternatively, it is possible to work with best $s-$sparse approximation of the vector. $l_p$-error of best $s$-term approximation of the coefficients, $\sigma_s(\mathbf{g})_p$, is defined as
\begin{equation}
\sigma_s(\mathbf{g})_p=\displaystyle\inf_{\mathbf z:\norm{\mathbf z}_0\leq s}\{\norm {\mathbf g-\mathbf z}_p \}.
\end{equation}
The goal is to recover the vector $\mathbf{g}$ or approximate it using samples of the function $g$. As we see in the section, one particular important feature for this purpose is an upper bound on Wigner-D functions of degree less that $B$, i.e. an upper bound on $\displaystyle\sup_{0 \leq l \leq B-1, -l \leq k,n \leq l}\norm{\mathrm D_l^{k,n}}_\infty$. As it is clear from \eqref{def:WigD}, it boils down to finding an upper bound for Wigner-d functions. We have seen that Wigner-d functions is nothing but weighted Jacobi polynomial. An upper bound on general weighted orthonormal functions is discussed in \cite[Theorem 6.1]{rauhut2012sparse} and also in \cite{szeg1939orthogonal}. However, we use directly the upper bound on Wigner-d function obtained in \cite[Theorem 1.1]{haagerup2014inequalities}.

\begin{lemma} [Bound for Jacobi polynomials Wigner-d functions \cite{haagerup2014inequalities}] \label{wignersmallbound}
For Jacobi polynomials $P_{\alpha}^{(\mu,\lambda)}$ of degree $\alpha$ and of order $(\mu,\lambda) $, there exists a constant $C\geq 0$ such that:
\begin{align}
\norm{(\sin \theta)^{1/2}  \sqrt{\gamma} \sin^{\mu}\bigg(\frac{\theta}{2}\bigg)& \cos^{\lambda}\bigg(\frac{\theta}{2}\bigg) P_{\alpha}^{(\mu,\lambda)}(\cos \theta)}_\infty\nonumber\\
&\leq C(2\alpha+\mu+\lambda+1)^{-1/4}.
\end{align} 
\end{lemma}
\begin{corollary} [ Bound for Wigner-d functions ] For Wigner-d function $\mathrm{d}_l^{k,n}(\cos \theta)$, there exists a constant $C\geq 0$ such that:
 $$
 \norm{(\sin \theta)^{1/2} \mathrm d_l^{k,n}(\cos \theta)}_\infty \leq C(2l +1)^{-1/4}.
 $$
\label{corol:upperbound}
\end{corollary}
The previous corollary is easily obtained using $\mu, \lambda \geq 0 $ defined as in Definition \ref{def:1} and observing that $2\alpha+\mu+\lambda$ equals $2l$. We will later use this corollary to find an upper bound on  weighted Wigner-D functions.

\subsection{Sparse Recovery for \ac{BOS} }
In this part, the main theorems concerning sparse recovery of \ac{BOS} are presented. We do not go through the details since the results are well known \cite{foucart2013mathematical,rauhut2012sparse}. A sufficient condition for sparse recovery is obtained in terms of \ac{RIP}. 
\begin{definition}
 The restricted isometry constant $\delta_s$ associated to the matrix $\mathbf A$ is the smallest number $\delta$ such that for all $s-$sparse vectors $\mathbf x$ 
$$
(1-\delta)\norm{\mathbf x}_2^2\leq \norm{\mathbf{Ax}}_2^2\leq (1+\delta)\norm{\mathbf x}_2^2.
$$
\end{definition}
It can be shown that a matrix that is constructed using random samples of orthonormal functions satisfy \ac{RIP}. Following theorem states this result. 
\begin{theorem}[\ac{RIP} for \ac{BOS} {\cite{foucart2013mathematical}}]\label{thm:RIP_BOS}
Consider a set of bounded orthonormal  basis  $\psi_j ,j \in [N]$ that are orthonormal with respect to measure $\nu$ on measurable space $\mathcal{D}$. Consider the matrix $ \boldsymbol{\psi} \in \mathbb{C}^{m \times N}$ with entries
\begin{equation*}
\boldsymbol{\psi}_{i,j} = \psi_j(t_i),\,\, i \in [m]\,\,,j \in [N]  
\end{equation*}
constructed with i.i.d. samples $t_i$ from the measure $\nu$. Suppose that $\sup_{j\in [N]}\norm{\psi_j}_\infty\leq {K}$. If 
\begin{equation*}
m \geq C\, \delta^{-2}\, {K}^2 \,s\, \log^3(s) \,\log(N)
\end{equation*}
then with probability at least $1-N^{-\gamma log^3(s)}$, the restricted isometry constant $\delta_s$ of $\frac{1}{\sqrt{m}} \psi$ satisfies $\delta_s \leq \delta$. The constants $C,\gamma \geq 0$ are universal.
\end{theorem}
Once the \ac{RIP} property is satisfied by a matrix, $s-$sparse vectors are recovered uniquely using \ac{BP} algorithm. intuitively, \ac{RIP} property implies null space property which is necessary and sufficient condition for unique recovery. Moreover \ac{RIP} property guarantees robust and stable recovery as shown in the following theorem.

\begin{theorem}[Sparse Recovery for \ac{RIP} Matrices {\cite{foucart2013mathematical}}]\label{thm:RIP_recovery}
Let the matrix $ \boldsymbol{\psi} \in \mathbb{C}^{m \times N}$ has restricted isometry constant $\delta_{2s}\leq 0.4931$. Suppose that the measurements are noisy $\mathbf y=\boldsymbol{\psi}\mathbf x+\boldsymbol{\eta}$ with $\norm{\boldsymbol{\eta}}_\infty\leq \epsilon$. If $\mathbf x^\#$ is the minimizer of  
\begin{equation*}
\arg\min \norm{\mathbf z}_1\text{ subject to } \norm{\mathbf y-\boldsymbol{\psi}\mathbf x}_2\leq \epsilon,
\end{equation*}
then $\norm{\mathbf x-\mathbf x^\#}_2\leq C_1\sigma_s(\mathbf x)_1+C_2\sqrt{s}\epsilon$ where $C_1,C_2$ depend only on $\delta_{2s}$. Without noise, we have $\mathbf x=\mathbf x^\#$ for $s-$sparse vectors $\mathbf x$.
\end{theorem}

\section{Main Results}\label{sec:Main}
Since Wigner-D functions are orthonormal, it suffices to find a useful upper bound $K$ on them and then using it in Theorem \ref{thm:RIP_BOS}.The following proposition serves this purpose.
\begin{proposition}[Bounds on preconditioned Wigner-D functions] \label{prep_wigner}
The Wigner-D functions $\mathrm D_l^{k,n}(\theta,\phi,\chi)$ preconditioned with $(\sin\theta)^{1/2}$
form orthonormal basis with respect to the product measure $\mathrm d\nu=\mathrm d\theta \mathrm d\phi \mathrm d\chi$ and satisfy the following upper bound. 
\begin{equation*}
\begin{aligned}
& \underset{\begin{subarray}{c}
   l,k,n 
  \end{subarray}}{\textnormal{sup}} 
& & \left\Vert (\sin\theta)^{1/2} \mathrm D_l^{k,n}(\theta,\phi,\chi) \right\Vert_{\infty}  \leq C_0 N^{\frac{1}{12}}
\end{aligned}
\end{equation*}
where $N$ is the total number of Wigner-D functions of degree less than $B$.
\end{proposition}
\begin{proof}
See Appendix
\end{proof} 

Note that the number of all orthonormal basis functions $N$ is related $B$ by $N =\frac{B(2B-1)(2B+1)}{3}$. The role of preconditioning function is to counter the increase of Wigner-D functions at the endpoints of the interval. Another bound can be obtained using \cite[Corollary 2]{burq2012weighted}. For a compact $n$-dimensional Riemannian manifold, a uniform bound on the first $N$ eigenfunctions are obtained as $N^{n-1/2n}$.  For $\mathrm{SO}(3)$, a 3-dimensional compact manifold, this approach yields the bound $N^{1/3}$ which is worse than the results above. As stated in \cite{burq2012weighted}, this bound deteriorates as the dimension of underlying manifold increases.  
Note that the general results in \cite{rauhut2012sparse,burq2012weighted} do not apply here since Wigner-D functions are not defined for surfaces of revolution. However the bound for Wigner-D functions is similar to the case of spherical harmonics in \cite{rauhut2011sparse} where $\norm{(\sin \theta)^{1/2}\, Y_l^k(\theta,\phi)}_\infty \leq C(l+1)^{1/4}$. 
Burq et all \cite{burq2012weighted} improved the bound using another preconditioning function to $\norm{(\sin^2\theta\cos \theta)^{1/6} \, Y_l^k(\theta,\phi)}_\infty \leq C(l+1)^{1/6}$ with respect to the measure $\mathrm d\nu=|\tan \theta|^{1/3}\mathrm  d\theta \mathrm d\phi$. In the numerical results, we also consider the performance of this measure. However, it is not clear at the moment how a similar bound can be obtained for eigenfunctions on $\mathrm{SO}(3)$.

From Proposition \ref{prep_wigner}, we can use Theorem \ref{thm:RIP_BOS} and \ref{thm:RIP_recovery} to prove sparse recovery guarantees for the coefficients of Wigner-D expansion using random samples of the function. The following theorem summarizes this result.

\begin{theorem}[RIP-BOS for Wigner-D functions]\label{thm:theorem_wigner}
Consider Wigner-D basis functions $\mathrm D_l^{k,n}(\theta,\phi,\chi)$ of degree less than $B$. Let $N$ be the number of these basis functions. Let the matrix $ \mathbf{A} \in \mathbb{C}^{m \times N}$ be such that the entries of row $i$ are $\mathrm D_l^{k,n}(\theta_i,\phi_i,\chi_i)$  where  $(\theta_i,\phi_i,\chi_i)$ are  i.i.d. samples using the product measure. Let $\mathbf{P}$ be a diagonal matrix with $P_{ii}= \sin(\theta_i)^{1/2}$. Suppose that the number of measurements satisfy the following inequality
\begin{equation*}
m \geq C^*\, {N}^{1/6} \,s\, \log^3(s) \,\log(N).
\end{equation*}
Also suppose that we observe the noisy measurements $\mathbf y=\mathbf A\mathbf g+\boldsymbol{\eta}$ with $\parallel \eta \parallel_{\infty} \leq \epsilon$. Then with  probability at least $1-N^{-\gamma log^3(s)}$, the following holds. If $\mathbf{g}^\#$ is the  solution to the following problem
\begin{equation*}
 \mathbf{g}^\# = \arg\min \left\Vert \mathbf{z} \right\Vert_1 \textnormal{subject to} \left\Vert \mathbf{P}\mathbf{A} \mathbf{z}- \mathbf{P}\mathbf{y}\right\Vert_2 \leq \sqrt{m}\epsilon. 
\end{equation*}
then, 
\begin{equation*}
 \left\Vert \mathbf {g} -\mathbf{g}^{\#}\right\Vert_2 \leq \frac{C_1 \sigma_s (\mathbf {g})_1}{\sqrt{s}} + C_2 \epsilon
\end{equation*}
The constant $C^*,C_1,C_2$ are universal. Without noise, the recovery is unique for $s-$sparse signal, namely $\mathbf {g}=\mathbf{g}^{\#}$.
\end{theorem}
\begin{proof}
See Appendix
\end{proof}

\section{Numerical Example}
\label{sec:Num}

\subsection{Simulation} 
The sparse recovery performance is studied for Wigner-D functions of degree less than $B=5$. For sparse recovery algorithm, we will use $\ell_1$-norm minimization package YALL1 \cite{zhang2010yall1}.  
 \begin{figure}[tb]
  \centering
   \includegraphics[width=0.9\textwidth]{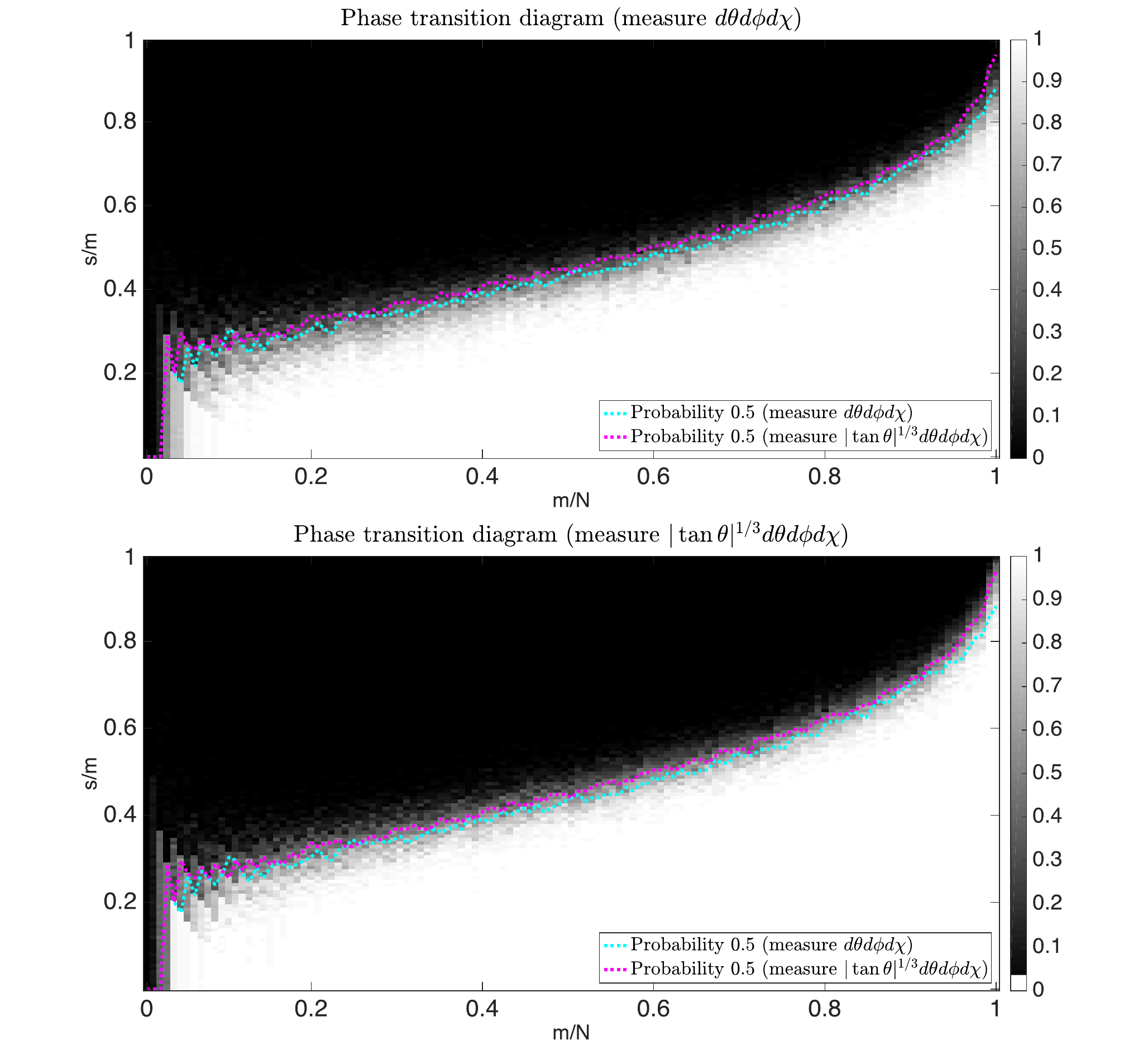}
   \caption{Sample measure (up :$d\theta d\phi d\chi$, down :$|\tan \theta|^{1/3}d\theta d\phi d\chi$)}
   \vspace{-2mm}
    \label{PDT}
\end{figure}
Figure \ref{PDT} shows the phase transition diagram for sparse recovery of Wigner-D basis expansion using random sampling by two measures, product measure $d\theta d\phi d\chi$ and the measure $|\tan \theta|^{1/3}d\theta d\phi d\chi$. The patch color around each point represents the recovery probability. The lines represent the probability of success around 0.5 for both measures. In this numerical example, we consider uniformly distributed support selection with standard normal distribution of non-zero values for the sparse vector. For each simulation, we count the frequency of success of $\ell_1$-norm minimization out of 50 trials with threshold $1e^{-3}$. It can be seen that measure $|\tan \theta|^{1/3}d\theta d\phi d\chi$ can improve the measurement bound compared to $d\theta d\phi d\chi$. 

\subsection{Spherical Near-Field Measurement}
\begin{figure}[tb]
  \centering
   \includegraphics[width=0.6\textwidth]{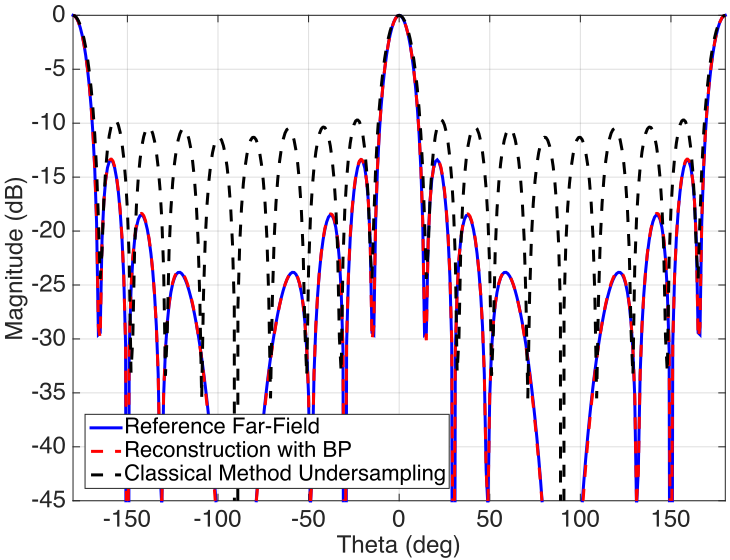}
   \caption{Far-field pattern of array 8 dipoles ($\phi$-cut= $0^\circ$)}
   \vspace{-2mm}
    \label{Farfield}
\end{figure}

The spherical near-field measurement of the antenna is defined in \cite{hansen1988spherical} and it uses the transmission formula as follows 
\begin{equation}
y(\theta,\phi,\chi)=v \sum \limits_{n=-v_{max}}^{v_{max}}\sum\limits_{h=1}^2\sum\limits_{l=1}^B\sum\limits_{k=-l}^l T_{hlk} \mathrm D_l^{k,n}(\theta,\phi,\chi)
\end{equation} 
where $y(\theta,\phi,\chi)$ is a bandlimited near-field sample with Wigner-D functions as basis, $h$ denotes the both transverse electric (TE) and magnetic (TM), $n$ and $\chi$ denote order and angle to measure polarization, respectively. Normally, it is desirable to measure co- and cross-polarization of the antenna and to use $n= \pm 1$, with angle $\chi \in \{0,\pi/2\}$. The goal is to estimate the transmission coefficient of the target antenna $T_{hlk}$ in near-field measurements and use it to determine far-field pattern. 
The classical method \cite{hansen1988spherical} uses Fourier analysis with equiangular sample to get the transmission coefficient $T_{hlk}$ and lacks the degree of freedom to choose different sampling pattern. In real measurement, we have to consider long duration of measurement time due to total number of samples even though the transmission coefficient is compressible and its support is smaller than measured total samples. In order to get better understanding of spherical near-field measurement we refer to \cite{hansen1988spherical,cornelius2015investigation}. Figure \ref{Farfield} shows far-field reconstruction after estimating transmission coefficient using $\ell_1$-norm minimization. It can be seen that the classical method fails to determine far-field pattern using same number of measurements as previously defined $\ell_1$ minimization.

\section{Conclusion and Discussion}\label{sec:Concl}
In this work, sparse recovery guarantees are provided for recovering sparse expansion in terms of Wigner-D functions. Using a bound on Wigner-d functions and Jacobi polynomials, an upper bound is obtained for Wigner-D functions which leads to constructing a measurement matrix satisfying \ac{RIP}. Numerically, phase transition diagram shows recovery condition for this basis function and shows how changing the sampling measure can improve the performance. 
However, it is interesting to note that if one could apply the bounds in \cite{burq2012weighted}, the measurement number would scale with $O(N^{1/9}s\log^3(s)\log(N))$. This possibility is verified by the numerical results using the measure $\mathrm d\nu=|\tan \theta|^{1/3}\mathrm  d\theta \mathrm  d\phi \mathrm  d\chi$. 
 Given the recent progress in \cite{klartag_improved_2014,nelson2014new,bandeira_discrete_2015}, another line of work consists in improving the \ac{RIP} bounds for \ac{BOS}. 
\section{Acknowledgement}
This work has been funded by DFG project- CoSStra-MA1184$|$31-1.
\vspace{-6mm}
\appendix
\section{Appendix}\label{sec:App}
\subsection{Proof of Proposition \ref{prep_wigner}}

Using Corollary \ref{corol:upperbound}, we can see that :
\begin{align*}
\norm{(\sin \theta)^{1/2} N_l &\,\mathrm D_l^{k,n}(\theta,\phi,\chi)}_\infty  = \norm{(\sin \theta)^{1/2} N_l\,d_l^{k,n}(\cos \theta)}_\infty\\
& \leq C\,N_l\,(2l +1)^{-1/4} = \frac{C}{\sqrt{8\pi^2}}(2l+1)^{1/4}\\
&\leq  \frac{C}{\sqrt{8\pi^2}}(2B-1)^{1/4}
\end{align*}
Note that the number of all orthonormal basis functions $N$ is related $B$ by $N =\frac{B(2B-1)(2B+1)}{3}$. Using  the inequality $(2B-1)^3\leq 6N$, we have for some constant $C_0$:
\begin{align*}
 \norm{(\sin \theta)^{1/2} N_l \,\mathrm D_l^{k,n}(\theta,\phi,\chi)}_\infty &\leq  \frac{C}{\sqrt{8\pi^2}}(6N)^{1/12}=C_0 N^{1/12}.
\end{align*}
\subsection{Proof of Theorem \ref{thm:theorem_wigner}}

Consider the functions $\varphi_l^{k,n}(\theta,\phi,\chi)= P(\theta)D_l^{k,n}(\theta,\phi,\chi)$, with product measure $\mathrm d\nu$. Note that the product measure $\mathrm d\nu=\mathrm  d\theta \mathrm  d\phi \mathrm d\chi$ with preconditioning function $P(\theta)^2=\sin(\theta)$ yields the uniform measure. Orthonormality can then be checked easily:
\begin{align*}
&\int_{\mathrm{SO}(3)}\varphi_l^{k,n}(\theta,\phi,\chi)\overline{\varphi_{l'}^{k',n'}\big(\theta,\phi,\chi\big)}\mathrm d\nu\\
&\int_{\mathrm{SO}(3)}\mathrm D_l^{k,n}(\theta,\phi,\chi)\overline{\mathrm D_{l'}^{k',n'}\big(\theta,\phi,\chi\big)}\sin(\theta)\mathrm  d\theta \mathrm  d\phi \mathrm d\chi=\delta_{nn'}\delta_{kk'}\delta_{ll'}.
\end{align*}
Therefore the functions $\varphi_l^{k,n}(\theta,\phi,\chi)$ form an orthonormal basis with bound provided in the Proposition \ref{prep_wigner}. Using these facts along with Theorem \ref{thm:RIP_BOS} and \ref{thm:RIP_recovery} finishes the proof. 

\vfill\pagebreak
\bibliographystyle{IEEEbib}
\bibliography{collection}
\end{document}